\documentclass[runningheads]{llncs}
\usepackage[english]{babel}
\usepackage[utf8]{inputenc}

\usepackage{url}
\usepackage{amsmath, amsfonts}

\usepackage{eqparbox}
\usepackage[T1]{fontenc}

\usepackage{graphicx}
\usepackage{subcaption}
\usepackage{color}
\usepackage{xcolor}
\usepackage{float}
\usepackage{multirow}
\usepackage{diagbox}
\usepackage[switch]{lineno}
\usepackage{makecell}

\usepackage[]{todonotes}
\usepackage[noend]{algpseudocode}
\usepackage[]{algorithm}

\usepackage[font=small,labelfont=bf]{caption}
\usepackage{wrapfig}

\usepackage{booktabs}
\usepackage{hyperref}
\usepackage{tikz}
\usetikzlibrary{arrows,automata,positioning}
\tikzstyle{accept}=[fill=gray]
\tikzset{initial text={}}
\usepackage{lipsum,adjustbox}

\usepackage{enumitem}
\setitemize{noitemsep,topsep=0pt,parsep=0pt,partopsep=0pt}
\setenumerate{noitemsep,topsep=0pt,parsep=0pt,partopsep=0pt}

\newcommand{\defas}[0]{\ensuremath{\stackrel{\mathrm{\scriptscriptstyle{def}}}{=}}}
\newcommand{\nat}[0]{\ensuremath{\mathbb{N}}}
\newcommand{\natpos}[0]{\ensuremath{\mathbb{N} \setminus \{ 0 \}}}
\newcommand{\real}[0]{\ensuremath{\mathbb{R}}}
\newcommand{\intinterval}[2]{\ensuremath{[{#1} \cdots {#2}]}}
\newcommand{\realinterval}[2]{[{#1}, {#2}]}
\newcommand{\vvector}[1]{\mathbf{#1}}
\renewcommand{\algorithmicwhile}{\textbf{while}}

\renewcommand{\algorithmicend}{\textbf{end}}

\makeatletter
\newcommand{\algmargin}{\the\ALG@thistlm}
\makeatother
\newlength{\whilewidth}
\algdef{SE}[parWHILE]{parWhile}{EndparWhile}[1]
  {\parbox[t]{\dimexpr\linewidth-\algmargin}{%
     \hangindent\whilewidth\strut\algorithmicwhile\ #1\ \algorithmicdo\strut}}{\algorithmicend\ \algorithmicwhile}%
\algnewcommand{\parState}[1]{\State%
  \parbox[t]{\dimexpr\linewidth-\algmargin}{\strut #1\strut}}

\usepackage{graphicx}
\usepackage{comment}
\usepackage{wrapfig}

\usepackage{amsmath}

\newcommand{\sig}[1]{{\textsf{{#1}}}}
\algnewcommand{\IIf}[1]{\State\algorithmicif\ #1\ \algorithmicthen}
\algnewcommand{\IElse}{\State\algorithmicelse\ }
\algnewcommand{\EndIIf}{\unskip }
\newcommand{\Comments}[1]{}

\begin{document}
\title{Prioritizing Corners in OoD Detectors via Symbolic String Manipulation}
\author{\hspace{-6mm}  Chih-Hong Cheng\inst{1} \and Changshun~Wu\inst{2} \and Emmanouil Seferis\inst{1} 
 \and Saddek Bensalem\inst{2}\thanks{The first two authors contributed equally to this work.}} 
\authorrunning{C.-H.~Cheng et al.}
\institute{
Fraunhofer IKS, Munich, Germany \\
\email{\{chih-hong.cheng,emmanouil.seferis\}@iks.frauhofer.de} \and
Univ. Grenoble Alpes, Verimag, Grenoble, France \\
\email{\{changshun.wu,saddek.bensalem\}@univ-grenoble-alpes.fr} 
}

\maketitle  % typeset the header of the contribution
\begin{abstract}
For safety assurance of deep neural networks (DNNs), out-of-distribution (OoD) monitoring techniques are essential as they filter spurious input  that is distant from the training dataset. This paper studies the problem of systematically testing  OoD monitors to avoid cases where an input data point is tested as in-distribution by the monitor, but the DNN produces spurious output predictions. We consider the definition of ``in-distribution'' characterized in the feature space by a union of hyperrectangles learned from the training dataset. Thus the testing is reduced to finding corners in hyperrectangles distant from the available training data in the feature space. Concretely, we encode the abstract location of every data point as a finite-length binary string, and the union of all binary strings is stored compactly using binary decision diagrams (BDDs). We demonstrate how to use BDDs to symbolically extract corners distant from all data points within the training set. Apart from test case generation, we explain how to use the proposed corners to fine-tune the DNN to ensure that it does not predict overly confidently. The result is evaluated over examples such as number and traffic sign recognition.
\end{abstract}
\section{Introduction}

To cope with practical concerns in autonomous driving where  deep neural networks (DNN)~\cite{goodfellow2016deep} are operated in an open environment, out-of-distribution (OoD) monitoring is a commonly used technique that raises a warning if a DNN receives an input distant from the training dataset. One of the weaknesses with OoD detection is regarding inputs that fall in the OoD detector's decision boundary while being distant from the training dataset. These inputs are considered ``in-distribution'' by the OoD detector but can impose safety issues due to extensive extrapolation. In this paper, we are thus addressing this issue by developing a disciplined method to identify the weakness of OoD detectors and improve the system accordingly. 

Precisely, we consider OoD detectors constructed using boxed abstraction-based approaches~\cite{henzinger2019outside,cheng2020towards,wu2021customizable}, where DNN-generated feature vectors from the training dataset are clustered and enclosed using hyperrectangles. The OoD detector raises a warning over an input, provided that its corresponding feature vector falls outside the boxed abstraction. We focus on analyzing the corners of the monitor's hyperrectangle and differentiate whether a corner is \emph{supported} or \emph{unsupported} depending on having some input in the training dataset generating feature vectors located in the corner. However, the number of \emph{exponentially many corners} in the abstraction reveals two challenges, namely (1) how to enumerate the unsupported corners and (2) how to prioritize unsupported corners to be analyzed.

\begin{itemize}
    \item For (1), we present an encoding technique that, for each feature vector dimension, decides if an input falls in the border subject to a closeness threshold~$\delta$. This allows encoding for each input in-sample as a binary string and storing the complete set compactly via Binary decision diagrams (BDDs)~\cite{bryant1992symbolic}. With an encoding via BDD, one can compute all unsupported corners using set difference operations.

\item For (2), we further present an algorithm manipulated on the BDDs that allows filtering all corners that are far from all training data subject to a minimum constant Hamming distance (which may be further translated into Euclidean distance). This forms the basis of our corner prioritization technique for abstractions characterized by a single hyperrectangle. For multiple boxed-abstraction, we use a lazy approach to omit the corners when the proposed corner from one box falls inside another box.

\end{itemize}

With a given corner proposal, we further encounter practical problems to produce input images that resemble ``natural'' images. We thus consider an alternative approach: it is feasible to have the DNN generate a prediction with low confidence for any input whose feature vectors resemble unsupported corners. This requirement leads to a DNN fine-tuning scheme as the final contribution of this paper:  The fine-tuning freezes parameters for all network layers before the monitored layer, thereby keeping the validity of the OoD monitor. However, it allows all layers after the monitored features to be adjusted. Thus the algorithm feeds the unsupported corners to the fine-tunable sub-network to ensure that the modified DNN reports every class with low confidence, while keeping the same prediction for existing training data.

We have evaluated our proposed techniques in applications ranging  from standard digit recognition to traffic sign detection. For corners inside the monitor while distant from the training data, our experiment indicates that the DNN indeed acts over-confidently in the corresponding prediction, which is later adjusted with our local training method. Altogether the positive evaluation of the technique offers a rigorous paradigm to align DNN testing, OoD detection, and DNN repair for safety-critical systems.

The rest of the paper is structured as follows: After reviewing related work in Section~\ref{sec:related.work}, we present in Section~\ref{sec:preliminaries} the basic notation as well as a concise definition on abstraction-based monitors. Subsequently, in Section~\ref{sec:test.case.proposal.single.box} we present our key results for prioritized corner case proposal in a single-box configuration and its extension to a multi-box setting. In Section~\ref{sec:applications}  we present how to use the discovered corners in improving the DNN via local training. Finally, we present our preliminary evaluation in Section~\ref{sec:evaluation} and conclude in Section~\ref{sec:concluding.remarks}.

\section{Related Work}\label{sec:related.work}

Systematically testing of DNNs has been an active research scheme, where readers may reference Section~5.1 of a recent survey~\cite{huang2020survey} for an overview of existing results. Overall, the line of attack is by first defining a coverage criterion, followed by concrete test case generation utilizing techniques such as adversarial perturbation~\cite{szegedy2013intriguing}, constraint solving~\cite{katz2017reluplex}, or model-based exploration~\cite{riccio2020model}. 
For white box coverage criteria, neuron coverage~\cite{pei2017deepxplore} and extensions (e.g., SS-coverage~\cite{sun2019structural} or neuron combinatorial testing~\cite{ma2018combinatorial}) essentially consider the activation pattern for neurons and demands the set of test inputs to satisfy a pre-defined relative completeness criterion; the idea is essentially motivated by classical software testing coverage (e.g., branch coverage) as used in safety standards. For black-box coverage criteria, multiple results are utilizing combinatorial testing~\cite{cheng2018quantitative,abrecht2021testing}, where by first defining the human-specified features in the input space, it is also possible to argue the relative completeness of the test data. For the above metrics, one can apply coverage-driven testing, i.e., generate test cases that maximally increase coverage. 
Note that the above test metrics and the associated test case prioritization techniques are not \emph{property-oriented}, i.e., prioritizing the test cases
does not have a direct relation with dependability attributes. 
This is in contrast to our work on testing the decision boundary of a DNN monitor, where our test prioritization scheme prefers corners (of the monitor) that have no input data being close-by. These corners refer to regions where DNN decisions are largely extrapolated, and it is important to ensure that inputs that may lead to these corners are properly tested.
The second differentiation is that we also consider the subsequent DNN repair scheme (via local training) to incorporate the distant-yet-uncovered corners.

In this paper, we are interested in testing the monitors built from an abstraction of feature vectors from the training data, where the shape of the abstraction is a union of hyperrectangles~\cite{henzinger2019outside,cheng2020towards,wu2021customizable}. There exist also other types of monitors. The most typical runtime monitoring approach for DNNs is to build a logic on top of the DNN, where the logic inspects some of the DNN features and tries to access the decision quality. Popular approaches in this direction are the baseline of Hendrycks et al.~\cite{hendrycks2016baseline} that looks at the output softmax value and flags it as problematic if lower than a threshold, or the ODIN approach that improves on it using temperature scaling~\cite{liang2017enhancing}. Further,~\cite{lee2018simple} looks at intermediate layers of a DNN and assumes that their features are approximately Gaussian-distributed. With that, they use the Mahanalobis distance as a confidence score for adversarial or OoD detection. The work of~\cite{lee2018simple}  is considered the practical state-of-the-art in the domain. In another direction, researchers have attempted to measure the uncertainty of a DNN on its decisions, using Bayesian approaches such as drop out at runtime~\cite{gal2016dropout} and ensemble learning. Deep Ensembles~\cite{lakshminarayanan2016simple} achieve state-of-the-art uncertainty estimation but at a large computational overhead (since one needs to train many models), thus recent work attempts to mitigate this with various ideas~\cite{dusenberry2020efficient,havasi2020training}. Although the above results surely have their benefits, for complex monitoring techniques, the decision boundary is never a single value but rather a complex geometric shape. For this, we observe a strong need in systematic testing over the decision boundaries (for rejecting an input or not), which is reflected in this work by testing or training against unsupported corners of a monitor.

\section{Preliminaries}

\label{sec:preliminaries}

Let $\nat$ and $\real$ be the sets of natural and real numbers.
To refer to integer intervals, we use $\intinterval{a}{b}$ with $a, b \in \nat$ and $a \leq b$.
To refer to real intervals, we use $\realinterval{a}{b}$ with $a, b \in \real \cup \{{-\infty}, \infty\}$ and if $a, b \in \real$, then $a \leq b$. We use square bracket when both sides are included, and use round bracket to exclude end points (e.g., $[a, b)$ for excluding~$b$). 
For $n \in \natpos$, $\real^n \defas \underbrace{ \real \times \cdots \times \real}_{n\ \text{times}}$ is the space of real coordinates of dimension $n$ and its elements are called $n$-dimensional vectors.
We use $\vvector{x} = (x_1, \ldots, x_n)$ to denote an $n$-dimensional vector.

\paragraph{Feedforward Neural Networks.}
A neuron is an elementary mathematical function.
A \emph{(forward) neural network}~$f \defas (g^{L}, \ldots , g^{1})$ is a sequential structure of $L \in \natpos$ layers, where, for $i \in \intinterval{1}{L}$, the $i$-th layer comprises $d_{i}$ neurons and implements a function $g^{i} : \mathbb{R}^{d_{i-1}} \rightarrow \mathbb{R}^{d_{i}}$.
The inputs of neurons at layer $i$ comprise (1) the outputs of neurons at layer $(i-1)$ and (2) a bias.
The outputs of neurons at layer~$i$ are inputs for neurons at layer~$i+1$.
Given a network input $\vvector{x} \in \real^{d_{0}}$, the output at the $i$-th layer is computed by the function composition
$f^{i}(\vvector{x}) \defas g^{i}(\cdots g^{2}(g^{1}(\vvector{x})))$. Therefore, $f^{L}(\vvector{x})$ is the output of the neural network. We use $f^{i}_j(\vvector{x})$ to extract the $j$-th value from the vector $f^{i}(\vvector{x})$.

\paragraph{Abstraction-based Monitors using Boxes~\cite{henzinger2019outside,cheng2020towards,wu2021customizable}.} In the following, we present the simplistic definition of abstraction-based monitors using multiple boxes. The definition is simplified in that we assume the monitor operates on all neurons within a given layer, but the technique is generic and can be used to monitor a subset of neurons across multiple layers. 

For a neural network~$f$ whose weights and bias related to neurons are fixed,  let $\mathcal{D}_{train} \defas \{(\vvector{x}, \vvector{y}) \;|\; \vvector{x} \in \mathbb{R}^{d_{0}}, \vvector{y} \in  \mathbb{R}^{d_{L}} \}$  be the corresponding training dataset.
We call $B \defas \big[ \realinterval{a_1}{b_1}, \cdots, \realinterval{a_n}{b_n} \big]$ an \textbf{$n$-dimensional box}, where $B$ is the set of points $\{(x_1, \ldots, x_n)\} \subseteq \real^n$ with $\forall i \in \intinterval{1}{n}: x_i \in \realinterval{a_i}{b_i}$.
Given a neural network $f$ and the corresponding training  dataset, let $k$ be a positive integer constant and $l \in \intinterval{1}{L}$. Then  $\mathcal{B}_{k,l,\delta} \defas \{B_1, \ldots, B_k\}$ is a \textbf{$k$-boxed abstraction monitor over layer~$l$ with buffer vector~$\delta \defas (\delta_1, \ldots, \delta_{d_l})$}, provided that $\mathcal{B}_{k,l,\vvector{\delta}}$ satisfies the following properties. 

\begin{enumerate}
    \item $\forall i \in \intinterval{1}{k}$, $B_i$ is a $d_{l}$-dimensional box. 
    
    \item $\forall (\vvector{x}, \vvector{y}) \in \mathcal{D}_{train}$, there exists $i \in \intinterval{1}{k}$ such that $ f^{l}(\vvector{x}) \in B_i$.
    
    \item $\forall i \in \intinterval{1}{k}$, let $B_i$ be $\big[ \realinterval{a_1}{b_1}, \cdots, \realinterval{a_{d_l}}{b_{d_l}} \big]$. Then 
    \begin{itemize}
        \item 
 for every $j \in \intinterval{1}{d_l}$, exists $(\vvector{x}, \vvector{y}) \in \mathcal{D}_{train}$ such that $ a_j  \leq  f^{l}_j(\vvector{x}) \leq a_j +\delta_j$, and 
  \item for every $j \in \intinterval{1}{d_l}$, exists $(\vvector{x'}, \vvector{y'}) \in \mathcal{D}_{train}$ such that $ b_j - \delta_j \leq  f^{l}_j(\vvector{x'}) \leq b_j$.
       \end{itemize}
\end{enumerate}

The three conditions stated above can be intuitively explained as follows: Condition~(1) ensures that any box is well formed, condition~(2) ensures that for any training data point, its feature vector at the $l$-th layer falls into one of the boxes, and (3) the construction of boxes is relatively tight in that for any dimension, there exists one training data point whose $j$-th dimension of its feature vector is close to (subject to $\delta_j$) the $j$-th lower-bound of the box; the same condition also holds for the $j$-th upper-bound. 

\paragraph{Monitoring.} Given a neural network $f$ and the boxed abstraction monitor $\mathcal{B}_{k,l,\delta}$, in runtime, the \textbf{monitor rejects an input $\vvector{x'}$} if $\not\exists i \in \intinterval{1}{k}: f^{l}(\vvector{x'}) \in B_i$. That is, the feature vector of $\vvector{x'}$ at the $l$-th layer is not contained by any box. As the containment checking $f^{l}(\vvector{x'}) \in B_i$  simply compares $f^{l}(\vvector{x'})$ against the box's lower and upper bounds on each dimension, it can be done in time linear to the number of neurons being monitored. 

\begin{wrapfigure}{R}{5cm}
\centering
\vspace{-10mm}
\includegraphics[width=.45\textwidth]{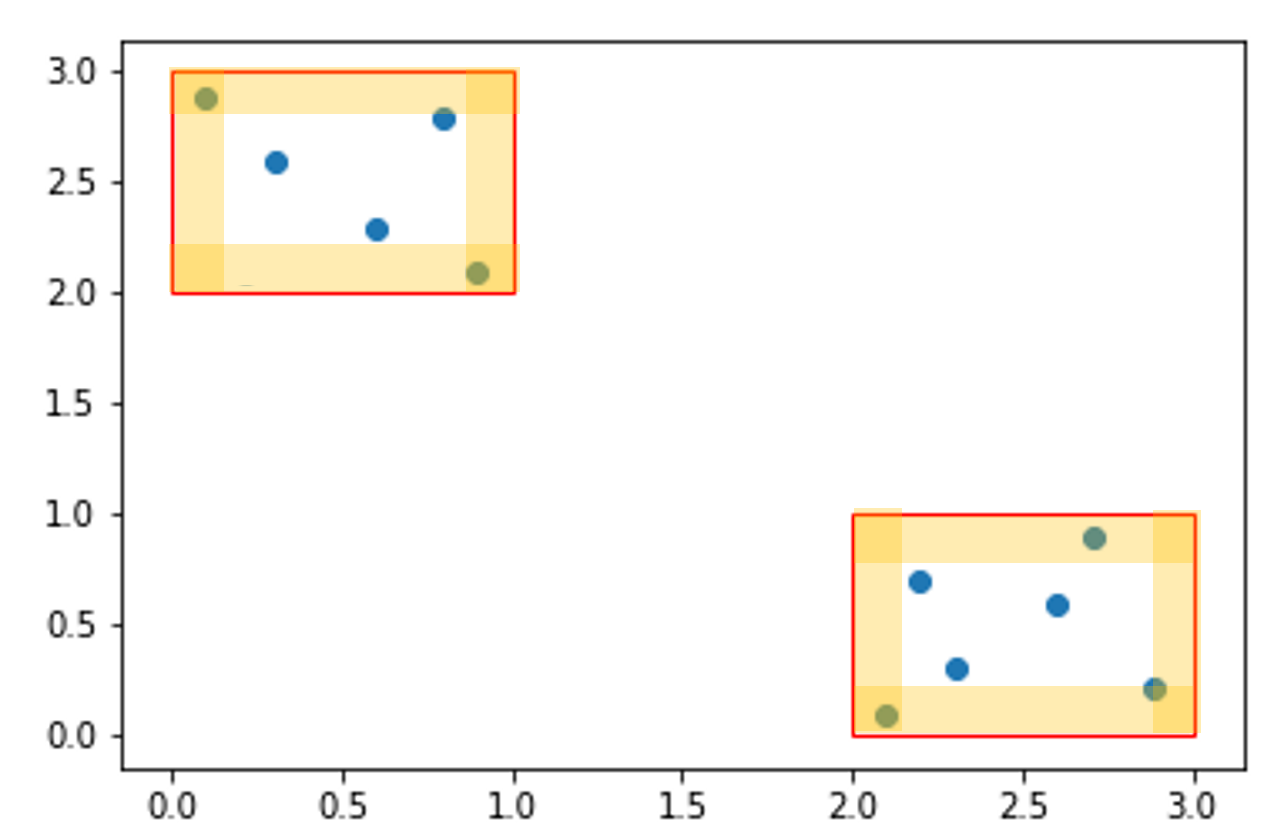}
  \caption{An example of two-boxes, where corners are deep-yellow areas. }\label{fig:2boxes}
\end{wrapfigure}

\begin{example}
Consider the set $\{ f^{l}(\vvector{x})\ |\ (\vvector{x}, \vvector{y}) \in \mathcal{D}_{train}\}  = \{ (0.1, 2.9), (0.3, 2.6),   (0.6, \\ 2.3), (0.8, 2.8), (0.9, 2.1), (2.1, 0.1), (2.2, 0.7), (2.3, 0.3), (2.6, 0.6), (2.9, 0.2), (2.7, 0.9) \} 
$ of feature vectors obtained at layer~$l$ that has only two neurons:
Fig.~\ref{fig:2boxes} shows \(
\mathcal{B}_{2, l, \delta} =
\{ \big[\realinterval{0}{1}, \realinterval{2}{3}\big], \big[\realinterval{2}{3},   \realinterval{0}{1}\big] \}
\), a $2$-boxed abstraction monitor with $\delta = (0.15, 0.15)$. The area influenced by $\delta$ is visualized in yellow.

\end{example}

\paragraph{Corners within monitors.}

As a monitor built from boxed abstraction only rejects an input if the feature vector falls outside the box, the borders of the box actually serve as a proxy for the boundary of the operational design domain (ODD) - anything inside a box is considered acceptable. With this concept in mind, we are interested in \textbf{finding test inputs that can lead to corners} of these boxes. As shown in Fig.~\ref{fig:2boxes}, for the box  $\big[\realinterval{0}{1}, \realinterval{2}{3}\big]$, the bottom left corner is not occupied by a feature vector produced from any training data point. 

We now precise the definition of corners. Given a box $B_i = \big[ \realinterval{a_1}{b_1}, \cdots, \realinterval{a_{d_l}}{b_{d_l}} \big] \in \mathcal{B}_{k,l,\delta}$, \textbf{the set of corners associated with $B_i$} is $C_{B_i} \defas \{\big[ \realinterval{\alpha_1}{\beta_1}, \cdots,  \realinterval{\alpha_{d_l}}{\beta_{d_l}} \big]\}$ where $\forall j \in \intinterval{1}{d_l}$, either

\begin{itemize}
    \item  $\realinterval{\alpha_j}{\beta_j} = \realinterval{a_j}{a_j + \delta_j}$, or 
    \item  $\realinterval{\alpha_j}{\beta_j} =  \realinterval{b_j -\delta_j}{b_j}$. 
\end{itemize}

\vspace{1mm}
Without surprise, the below lemma reminded us the well known problem of \emph{combinatorial explosion}, where the number of corners, although linear to the number of boxes, is exponential to the number of dimensions. 

\begin{lemma}\label{lemma:exponential}
Given $\mathcal{B}_{k,l,\delta}$,  $\sum^{k}_{1} |C_{B_i}|$, i.e., the total number of corners associated with the monitor, equals $k\cdot 2^{d_l}$. 
\end{lemma}

\begin{figure}[t]
    \centering
    \includegraphics[width=.8\columnwidth]{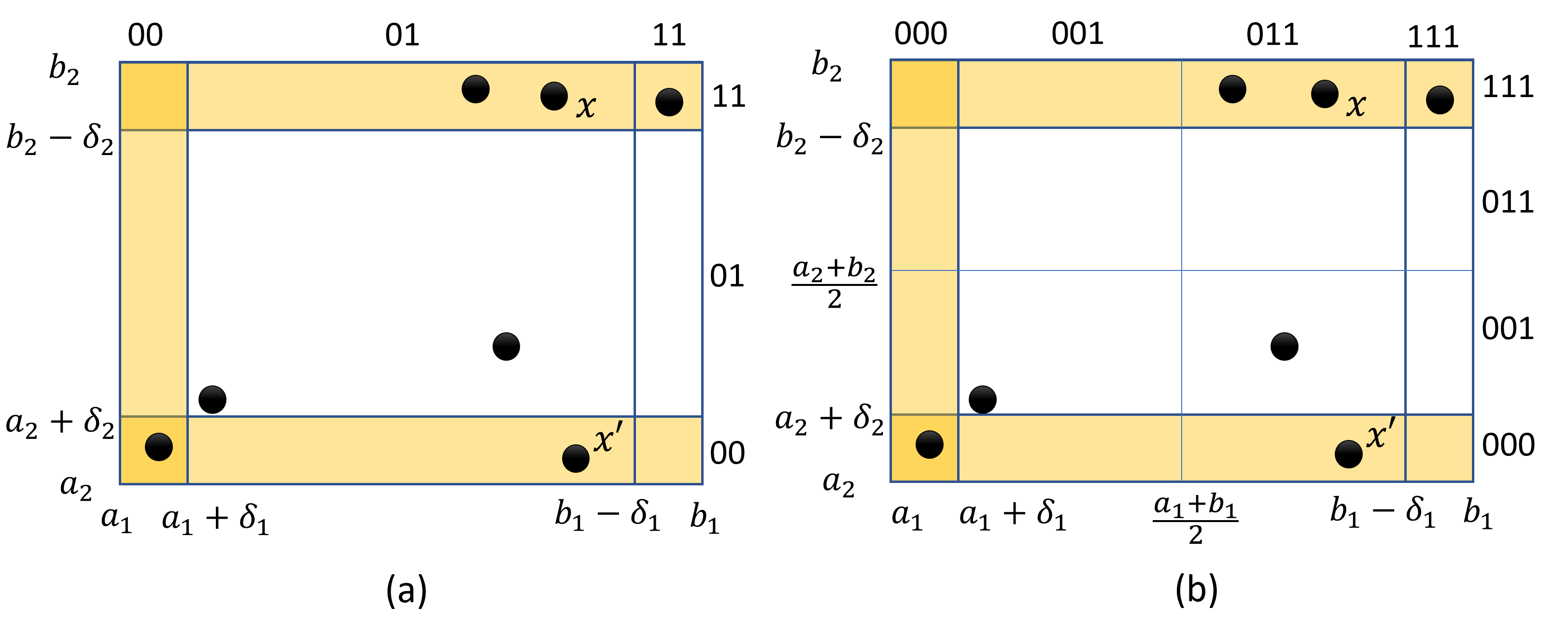}\caption{Partition the boxed monitor and encode every region using BDDs. A black dot represents a feature vector generated from a training data point.}
    \label{fig:kBitsEncoding}
    \vspace{-5mm}
\end{figure}

Given the set $C_{B_i}$ of corners associated with $B_i$, define  $C^{s}_{B_i} \subseteq C_{B_i}$ to be the \textbf{(training-data) supported corners} where for each $\big[ \realinterval{\alpha_1}{\beta_1}, \cdots,  \realinterval{\alpha_{d_l}}{\beta_{d_l}} \big]$ in $C^{s}_{B_i}$,  exists $(\vvector{x}, \vvector{y}) \in \mathcal{D}_{train}$ such that  $\forall j \in \intinterval{1}{d_l}:   f^{l}_j(\vvector{x}) \in \realinterval{\alpha_{j}}{\beta_{j}}$. The set of \textbf{(training-data) unsupported corners} $C^{u}_{B_i}$ is the set complement, i.e., $C^{u}_{B_i} \defas C_{B_i} \setminus C^{s}_{B_i}$. As an example, consider the box $B_i$ in Fig.~\ref{fig:kBitsEncoding}(a). The set of unsupported corners $C^{u}_{B_i}$ is $\{\big[[a_1, a_1 + \delta_1], [b_2 - \delta_2, b_2]], \big[[b_1 - \delta_1, b_1], [a_2, a_2 + \delta_2]]\}$, i.e, the top-left corner and the bottom-right corner.

An unsupported corner reflects the possibility of having an input~$\vvector{x}_{op}$ in operation time, where the DNN-computed $l^{th}$-layer feature vector $f^l(\vvector{x}_{op})$ falls into the corner of the monitor. It reflects additional risks, as we do not know the prediction result, but the monitor also will not reject the input. The consequence of Lemma~\ref{lemma:exponential} implies that when we only have a finite budget for testing unsupported corners, we need to develop methods to \textbf{prioritize} them, as detailed in the following sections.

\section{Unsupported Corner Prioritization under Single-Boxed Abstraction}\label{sec:test.case.proposal.single.box}

We first consider the special case where only one box is used in the monitoring. That is, we consider $\mathcal{B}_{1,l,\delta} = \{B\}$ where $B = \{(x_1, \ldots, x_{d_l}) \, | \,  x_1 \in \realinterval{a_1}{b_1}, \ldots, x_{d_l} \in \realinterval{a_{d_l}}{b_{d_l}}\}$. The workflow is to first consider encoding feature vectors at the $l$-th layer into fixed-length binary strings, in order to derive the set of unsupported corners. Subsequently, prioritize the unsupported corners via Hamming distance-based filtering. The algorithm stated in this section serves as the foundation for the general multi-boxed monitor setting detailed in later sections.  

\subsection{Encoding feature vectors using binary strings}\label{subsec:encoding.binary.string}

Given a finite-length Boolean string $\vvector{b} \in \{0, 1\}^{*}$, We use $\vvector{b}_{\intinterval{i}{j}}$ to denote the substring indexed from~$i$ to $j$. For a single-boxed monitor $\mathcal{B}_{1,l,\delta} = \{B\}$ constructed from $\mathcal{D}_{train}$, let the \textbf{$\phi$-bit encoding} ($\phi \geq 2$) be a function $\sig{enc}^{\phi}: \real^{d_l} \rightarrow \{0,1\}^{\phi\cdot {d_l}}$ that, for any $\vvector{x}\in \mathcal{D}_{train}$, translates the feature vector $f^l(\vvector{x})$ to a Boolean string~$\vvector{b}$ (with length $\phi\cdot{d_l}$) using the following operation: $\forall j \in \intinterval{1}{d_l}$, %$\vvector{b}_{\intinterval{\phi j}{\phi (j+1) -1}}$,

\vspace{2mm}

\begin{itemize}
    \item if 
    $f^l_j(\vvector{x}) \in [\alpha_j, \alpha_j + \delta_j]$, then 
    $\vvector{b}_{\intinterval{\phi (j-1) +1 }{\phi  j}} =\underbrace{0 \cdots 0}_{\phi\ \text{times}} $;
    
       \item  else if 
    $f^l_j(\vvector{x}) \in [\beta_j -\delta_j, \beta_j]$, then $\vvector{b}_{\intinterval{\phi (j-1) +1 }{\phi  j}} = \underbrace{1 \cdots 1}_{\phi\ \text{times}}$;

    \item otherwise,  $\vvector{b}_{\intinterval{\phi (j-1) +1 }{\phi j}} =\underbrace{0 \cdots 0}_{\phi-\tau\ \text{times}}\underbrace{1 \cdots 1}_{\tau\ \text{times}}$  when \\ $f^l_j(\vvector{x}) \in [a_j + \delta_j + \frac{(\tau - 1 )(b_j-a_j-2\delta_j)}{\phi \ -1}, a_j + \delta_j + \frac{(\tau)(b_j-a_j-2\delta_j)}{\phi \ -1})$
    
\end{itemize}

\vspace{2mm}

\noindent The $\phi$-bit encoding essentially considers $f^l(\vvector{x})$ in  dimension $j$, assigns the substring with all $0$s when $f^l_j(\vvector{x})$ falls in the corner reflecting the lower-bound, assigns with all $1$s when $f^l_j(\vvector{x})$ falls in the corner reflecting the upper-bound, and finally, splits the rest interval of length~$b_j-a_j-2\delta_j$ into~$\phi -1$ equally sized intervals and assigns each interval with an encoding. 
Fig.~\ref{fig:kBitsEncoding} illustrates the result of $2$-bit and $3$-bit partitioning under a 2-dimensional boxed monitor. For point~$\vvector{x}$ in Fig.~\ref{fig:kBitsEncoding}(b), $\sig{enc}^{3}(\vvector{x}) = 011111$. The first part ``$011$'' comes as when $\tau=2$, $f^l_1(\vvector{x}) \in [a_1 + \delta_1 + \frac{(2 - 1 )(b_1-a_1-2\delta_1)}{3 \ -1}, a_1 + \delta_1 + \frac{(2)(b_1-a_1-2\delta_1)}{3 \ -1})$.  The second part ``$111$'' comes as $f^l_2(\vvector{x}) \in [\beta_2 -\delta_2, \beta_2]$. Given an input $\vvector{x}$ and its computed feature vector $f^l(\vvector{x})$, the time required for perfotming $\phi$-bit encoding  is in low degree polynomial with respect to~$d_l$ and~$\phi$.

\subsection{BDD encoding and priortizing the unsupported corners}

This section presents Algorithm~\ref{alg:priortizing}, a BDD-based algorithm for identifying unsupported corners. To ease understanding, we separate the algorithm into three parts.  

\paragraph{A: Encode the complete training dataset.} Given the training dataset $\mathcal{D}_{train}$ and the DNN function $f$, one can easily compute $\{ \vvector{b} \ | \ \vvector{b}= \sig{enc}^{\phi}(f^l(\vvector{x})) \ \text{where} \ \vvector{x}\in  \mathcal{D}_{train}\}$ as be the set of all binary strings characterizing the complete training dataset. As each element in the set is a fixed-length binary string, the set can be compactly stored using Binary Decision Diagrams. 

Precisely, as the length of a binary string $\vvector{b}=\sig{enc}^{\phi}(f^l(\vvector{x}))$ equals $\phi \cdot d_l $, in our encoding we use $\phi \cdot d_l$ BDD variables, denoted as $\sig{bv}_1, \ldots, \sig{bv}_{\phi d_l}$, such that $\sig{bv}_i = \sig{true}\ \text{iff} \ \vvector{b}_{\intinterval{i}{i}} = 1$. Line~$1$ of Algorithm~\ref{alg:priortizing} performs such a declaration. Lines 2 to 9 perform the BDD encoding and creation of the set $S_{train}$ containing all binary strings created from the training set. Initially (line~2) $S_{train}$ is set to be an empty set.  Subsequently, generate the binary string (line~4), and encode  a set $S_{\vvector{b}}$ which contains only the binary string (line 5-8). Finally, add  $S_{\vvector{b}}$ to $S_{train}$ (line~9).

\begin{algorithm}[t]
\caption{Priortizing unsupported corners using BDD}\label{alg:priortizing}
\begin{algorithmic}[1]
\Require Dataset $\mathcal{D}_{train}$, DNN $f$, 1-box monitor $\mathcal{B}_{1,l,\delta} = \{B\}$, $\phi$, distance metric $\Delta$
\Ensure The set $\{\vvector{b}^u\}$ of binary strings represented in BDD, reflecting unsupported corners $C^{u}_{B}$ for box~$B$, with each $\vvector{b}^u$ distant to all training data encodings by at least $\Delta + 1$ bits.
\State Declare BDD variables $\sig{bv}_1, \ldots, \sig{bv}_{\phi d_l}$.
\State $S_{train} \gets \sig{BDD.false}$ \Comment{Initialize to empty set}
\ForAll{$\vvector{x} \in \mathcal{D}_{train}$}
    \State $\vvector{b} \gets \sig{enc}^{\phi}(f^l(\vvector{x}))$
    \State $S_{\vvector{b}}  \gets \sig{BDD.true}$
    \ForAll{$m \in \intinterval{1}{\phi d_l}$} \Comment{Refine the set to contain only $\vvector{b}$}
    \IIf{$\vvector{b}_{\intinterval{m}{m}} = 1$}           $S_{\vvector{b}}  \gets \sig{BDD.and}(S_{\vvector{b}}, \sig{bv}_m)$
    \EndIIf
   \IElse{}
    $S_{\vvector{b}}  \gets \sig{BDD.and}(S_{\vvector{b}}, \sig{BDD.not}(\sig{bv}_m))$
    \EndIIf
    \EndFor
    \State $S_{train} \gets \sig{BDD.or}(S_{train}, S_{b})$ \Comment{Add $S_{b}$ to the set}
\EndFor

\State $S_{all.corners} \gets \sig{BDD.true}$ 

\ForAll{$j \in \intinterval{1}{d_l}$}
\State $S_{j0s} \gets \sig{BDD.true}$; $S_{j1s} \gets \sig{BDD.true}$ 
\ForAll{$m \in \intinterval{1}{\phi}$}
\State $S_{j0s} \gets \sig{BDD.and}(S_{j0s}, \sig{BDD.not}(\sig{bv}_{\phi (j-1) + m}))$ 
\State $S_{j1s} \gets \sig{BDD.and}(S_{j1s}, \sig{bv}_{\phi (j-1) + m})$ 
\EndFor
\State $S_{all.corners} \gets \sig{BDD.and}(S_{all.corners}, \sig{BDD.or}(S_{j0s}, S_{j1s}))$ 
\EndFor

\State $S_{unsup} \gets S_{all.corners} \setminus S_{train}$ \Comment{$\sig{BDD.setminus}(\cdot , \cdot)$ operation}

\State  $S^{\leq \Delta}_{train} \gets S_{train}$

\ForAll{$n \in \intinterval{1}{\Delta}$} 

\State  $S_{local} \gets S^{\leq \Delta}_{train}$

\ForAll{$m \in \intinterval{1}{\phi d_l}$} 
\State $S^{\leq \Delta}_{train} \gets \sig{BDD.or}(S^{\leq \Delta}_{train}, \sig{BDD.exists}(S_{local}, \sig{bv}_{m}))$

\EndFor
\EndFor 

\State return $S_{unsup}  \setminus S^{\leq \Delta}_{train}$

\end{algorithmic}
\end{algorithm}

\paragraph{B: Derive the set of unsupported corners.} 

Lines 10 to 17 of Algorithm~\ref{alg:priortizing} computes $S_{unsup}$, where each binary string in $S_{unsup}$ corresponds to an unsupported corner. The set is computed by a set difference operation (line 17) between the set of all corners $S_{all.corners}$ and $S_{train}$. Following the encoding in Section~\ref{subsec:encoding.binary.string}, we know that the set of all corners corresponds to $\{\underbrace{0 \cdots 0}_{\phi\ \text{times}}, \underbrace{1 \cdots 1}_{\phi\ \text{times}}\}^{d_l}$. As an example, in Fig.~\ref{fig:kBitsEncoding}(b), the set of all corners equals $\{000000,  000111, 111000, 111111\}$. Lines 10 to 16 of Algorithm~\ref{alg:priortizing} describe how such a construction can be done symbolically using BDD, where the number of BDD operations being triggered is linear to $\phi \cdot d_l$. The set $S_{j0s}$, after the inner loop (line 13-15), contains the set of all possible Boolean words with the restriction that $\vvector{b}_{\intinterval{\phi (j-1) +1}{\phi j}}$ equals $\underbrace{0 \cdots 0}_{\phi\ \text{times}}$ (similarly $S_{j1s}$  for having 1s). The ``\sig{BDD.or}'' operation at line~16 performs a set union operation between  $S_{j0s}$  and  $S_{j1s}$, to explicitly allow two types of possibilities within $\vvector{b}_{\intinterval{\phi (j-1) +1}{\phi j}}$.

\paragraph{C: Filter unsupported corners that are close to training data.}

Although at line~17 of Algorithm~\ref{alg:priortizing}, all unsupported corners are stored compactly inside the BDD, the implication of Lemma~\ref{lemma:exponential} suggests that the number of unsupported corners can still be exponential. Therefore, we are interested in further filtering out some unsupported corners and only keeping those unsupported corners that are distant from the training data.

Consider again the example in Fig.~\ref{fig:kBitsEncoding}(b), where $S_{unsup}$ is the symbolic representation of two strings, namely
\begin{itemize}
    \item $000111$ reflecting the top-left corner, and 
     \item $111000$ reflecting the bottom-right corner. 
\end{itemize}
The algorithm thus should keep $000111$ and filter $111000$, as the bottom-right corner has a training data~$\vvector{x}'$ being close-by.

\vspace{2mm}

The final part of Algorithm~\ref{alg:priortizing} (starting at line~18) describes how to perform such an operation symbolically by utilizing the Hamming distance on the binary string level. Consider again the example in  Fig.~\ref{fig:kBitsEncoding}(b), where for training data~$\vvector{x}'$, $\sig{enc}^3(f^l(\vvector{x}')) = 011000$. The Hamming distance between ``$011000$'' and the bottom-right corner encoding ``$111000$'' equals~$1$. For the top-left corner having its encoding being $000111$, there exists only data points whose encoding (e.g., $\vvector{x}$ has an encoding of $011111$) has a Hamming distance of~$2$. Therefore, by filtering out the elements with Hamming distance $1$, only the top-left corner is kept.

Within Algorithm~\ref{alg:priortizing}, line~18 maintains $S^{\leq \Delta}_{train}$ as a BDD storing every binary string that has another binary string in $S_{train}$ such that the Hamming distance between these two is at most~$\Delta$. Initially, $S^{\leq \Delta}_{train}$ is set to be $S_{train}$, reflecting the case of Hamming distance being~$0$.  The loop of Line~19 is executed $\Delta$ times to gradually increase $S^{\leq \Delta}_{train}$ to cover strings with Hamming distance from~$1$ up to~$\Delta$. 

Within the loop, first a local copy $S_{local}$ is created (line 20). Subsequently, enlarging the set by a Hamming distance~1 can be done by the inner loop within line 21-22: for each variable index~$m$, perform existential quantification over the local copy to get the set of binary strings that is insensitive at variable $\sig{bv}_m$. As an example, if $S_{local} = \{011000\}$, then performing existential quantification on the first variable generates a set  ``$\{\theta 11000 \;|\; \theta \in {0,1}\}$'', and performing existential quantification on the second variable generates another set  ``$\{0\theta1000 \;|\; \theta \in {0,1}\}$''. A union over all these newly generated sets returns the set of strings whose Hamming distance to the original ``011000'' is less or equal to~$1$. 

Finally, line~23 performs another set difference to remove elements in $S_{unsup}$ that is present in $S^{\leq \Delta}_{train}$, and the resulting set is returned as the output of the algorithm.

\subsection{Corner prioritization with multi-boxed abstraction monitors}\label{subsec:test.case.proposal.multi.box}

In the previous section, we focus on finding corners within a box, where the corners are distant (by means of Hamming distance) to DNN-computed feature vectors from the training dataset. Nevertheless, when the monitor uses multiple boxes, is it possible that the corner being prioritized in one box has been covered by another box? An example can be found in Fig.~\ref{fig:2overlapping}, where the monitor contains two boxes $B_1$ and $B_2$. If the algorithm applied on $B_1$ proposes corner~$c_1$ to be tested, it would be a waste as~$c_1$ lies inside~$B_2$.

We propose a lazy approach to mediate this problem - whenever a corner proposal is created from one box, use a strengthened condition and check if some part of the corner is deep inside another box (subject to~$\delta$). Precisely, given $\mathcal{B}_{k,l,\vvector{\delta}}$, provided that Algorithm~\ref{alg:priortizing} applied on $B_i = \big[ \realinterval{a_1}{b_1}, \cdots, \realinterval{a_{d_l}}{b_{d_l}} \big] \in  \mathcal{B}_{k,l,\vvector{\delta}}$ suggests an unsupported corner $c \in C^{u}_{B_i}$ whose corresponding binary string equals  $\vvector{b}$, conduct the following:

\begin{wrapfigure}{R}{5cm}
\centering
  \vspace{-10mm}
  \includegraphics[width=.35\columnwidth]{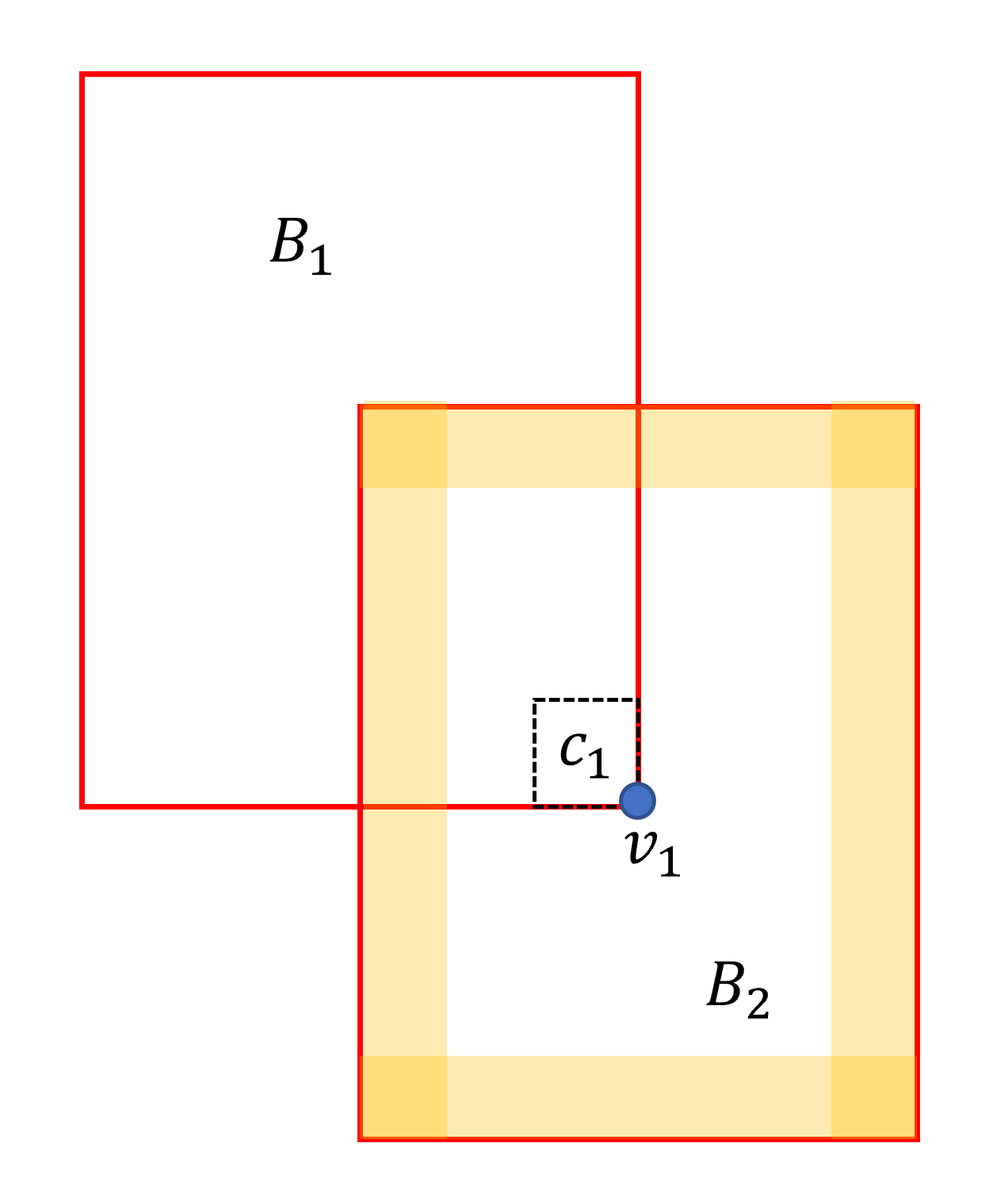}
  \caption{Two overlapping boxes. }\label{fig:2overlapping}
  \vspace{-15mm}
\end{wrapfigure}

\begin{enumerate}
    \item Given $\vvector{b}$, find a vertex  $\vvector{v} = (v_1, \ldots, v_{d_l})$ in  box $B_i$ that is also in the proposed corner~$c$. Precisely, for $\forall j \in \intinterval{1}{d_l}$, 
    \begin{itemize}
        \item if $\vvector{b}_{\intinterval{\phi(j-1) +1}{\phi j}} = \underbrace{0 \cdots 0}_{\phi\ \text{times}}$, set~$v_j$ to be~$a_j$.
        \item Otherwise, set~$v_j$ to be~$b_j$.
        
    \end{itemize}
    \item Discard the corner proposal on~$c$, whenever there exists  $B_{i'} = \big[ \realinterval{a'_1}{b'_1}, \cdots,\\  \realinterval{a'_{d_l}}{b'_{d_l}} \big] \in \mathcal{B}_{k,l,\vvector{\delta}}$, $i' \neq i$, such that the following holds: $\forall j \in \intinterval{1}{d_l}: a'_j + \delta_j < v_j< b'_j - \delta_j$.
\end{enumerate}

\vspace{2mm}
\noindent The time complexity for rejecting a corner proposal is in low degree polynomial: %$\mathcal{O}(k(d_l)^2)$.
\begin{itemize}
    \item For step (1), assigning each $v_j$ sums up the time $\mathcal{O}(d_l)$.
    
    \item  For step (2), the containment check is done on every other box (the number of boxes equals~$k$) over all dimensions (size $d_l$), leading to the time complexity $\mathcal{O}(k \cdot d_l)$. 
\end{itemize}

\section{Improving the DNN against the Unsupported Corners}\label{sec:applications}

As unsupported corners represent regions in the monitor where no training data is close-by, any input whose feature vector falls in that corner will not be rejected by the monitor, leading to safety concerns if the prediction is incorrect. For classification tasks, one possible mediation is  to explicitly ensure that any input whose feature vector falls in the unsupported corner does not cause the DNN to generate a strong prediction over a particular class. 

As an example, if the DNN~$f$ is used for digit recognition and $d_{L}$ equals~$10$ with each $f^{(L)}_i$ indicating the possibility of the character being~$i-1$, it is desirable to let an input~$\vvector{x}$, whose feature vector falls inside the unsupported corner, to produce $f^{(L)}_1(\vvector{x}) \cong f^{(L)}_2(\vvector{x}) \cong \ldots \cong f^{(L)}_{10}(\vvector{x}) \cong 0.1$, i.e., the DNN is not certain on which class this input belongs to. One can naively retrain the complete DNN against such an input~$\vvector{x}$. Nevertheless, if the  DNN is completely retrained, the created monitor $\mathcal{B}_{k, l, \delta}$ is no longer valid, as the parameters before layer~$l$ have been changed due to re-training.

\begin{algorithm}[t]
\caption{DNN modification against unsupported corners under 1-boxed abstraction monitor (classification network with one-hot output encoding)}\label{alg:modification}
\begin{algorithmic}[1]
\Require Dataset $\mathcal{D}_{train}$, DNN $f = (g^{L}, \ldots, g^{1})$, 1-boxed monitor $\mathcal{B}_{1,l,\delta}$, $S \subset S_{unsup}  \setminus S^{\leq \Delta}_{train}$ created from Algorithm~\ref{alg:priortizing}, the number of samples $\rho$ per unsupported corner. 
\Ensure Updated DNN $f'$.
\State Create dataset $\mathcal{D}_{modify} \defas \{(f^l(\vvector{x}), \vvector{y})) \ | \ (\vvector{x},\vvector{y}) \in \mathcal{D}_{train} \}$
%\ForAll{$\vvector{b} \in S_{unsup}  \setminus S^{\leq \Delta}_{train}$}
\ForAll{$\vvector{b} \in S$}
\parState{Let $\vvector{c} \defas \big[ \realinterval{\alpha_1}{\beta_1}, \cdots,  \realinterval{\alpha_{d_l}}{\beta_{d_l}} \big] \in \real^{d_l} $ be the corresponding corner of $\vvector{b}$.  }
\parState{Sample $\rho$ points $\vvector{p}_1, \ldots, \vvector{p}_{\rho}$ from $\vvector{c}$.  }
\ForAll{$i \in \intinterval{1}{\rho}$}
\State    $\mathcal{D}_{modify} \gets \mathcal{D}_{modify} \cup \{(\vvector{p}_i, (\frac{1}{d_L}, \ldots, \frac{1}{d_L}))\}$
\EndFor
\EndFor

\State{Improve $g^L, \ldots g^{l+1}$ to $\hat{g}^L, \ldots \hat{g}^{l+1}$ by training against $\mathcal{D}_{modify}$}

\State Return $f' \defas (\hat{g}^L, \ldots \hat{g}^{l+1}, g^l, \ldots, g^1)$

\end{algorithmic}
\end{algorithm}

Towards this issue, Algorithm~\ref{alg:modification} presents a local DNN modification scheme\footnote{For simplicity, we only show the algorithm for 1-boxed abstraction monitor, while extensions for multi-boxed abstraction monitor can follow the same paradigm stated in Section~\ref{subsec:test.case.proposal.multi.box}.}   where the re-training is only done between layers~$l+1$ and~$L$. As the new DNN share the same function with the existing one from layer $1$ to layer~$l$, previously constructed 1-boxed monitor remains applicable in the new DNN. 

As re-training is only done over a sub-network between layers~$l+1$ and~$L$, the input for training the sub-network  is the output of layer~$l$. Therefore, reflected at line~1, one prepares a new training dataset where the input is $f^{l}(\vvector{x})$. The input for Algorithm~\ref{alg:modification} also contains~$S$, which is a subset of unsupported corners derived from Algorithm~\ref{alg:priortizing}. Lines~2 to~6  translate each binary string in~$S$ into an unsupported corner (line~3) and sample~$\rho$ points (line~4) to be added to the new training dataset. As stated in the previous paragraph, we wish the result of these points to be unbiased for any output class. Therefore stated at line~6, the corresponding label, under the assumption where~$\mathcal{D}_{train}$ uses one-hot encoding, should be~$(\frac{1}{d_{L}}, \ldots, \frac{1}{d_{L}})$.

\section{Evaluation}\label{sec:evaluation}

This section aims to experimentally answer two questions about the unsupported corners generated by the method in Section \ref{sec:test.case.proposal.single.box}.
The first question is regarding the behavior of feature vectors in the unsupported corners reflected in the output (Section~\ref{subsec.post.evaluation}).  The second question is regarding generating  inputs that can lead to these unsupported corners (Section~\ref{subsec.test.case.gen}).

Specifically, we consider monitors built on the penultimate layer of two neural networks, trained on benchmarks MNIST \cite{lecun2010mnist} and GTSRB \cite{Houben-IJCNN-2013}, respectively, to classify handwritten digits (0-9) and traffic signs.
Following Algorithm~\ref{alg:priortizing}, we first encode the monitors' supported corners using BDD representation. Subsequently, compute the unsupported corners using symbolic set difference operations. 
We use Pytorch\footnote{\url{https://pytorch.org/}} to train the DNN and use the python-based BDD library dd\footnote{\url{https://github.com/tulip-control/dd}} for encoding the binary strings into the BDD.

\begin{table}[t]
\centering
\caption{Hyper-parameter setting in the experiments} 
\label{tab:hyperParemeters}
\resizebox{\textwidth}{!}{%
\begin{tabular}{|c|c|c|c|}
\hline
dataset & \# of monitored neurons & $m: $\#  unsupported corners & $\rho$: \# collected samples per corner \\ \hline
MNIST   & 40                      & 1000                            & 10                       \\ \hline
GTSRB   & 84                      & 1000                            & 10                       \\ \hline
\end{tabular}%
}
\end{table}

\vspace{-2mm}
\subsection{Understanding unsupported corners}\label{subsec.post.evaluation}

This subsection focuses on understanding the output softmax (probability) values for the feature vectors from unsupported corners.
We take~$m$ unsupported corners and from each of them uniformly pick~$\rho$ samples in the corresponding corner. The hyper-parameters used in the experiments are shown in Table \ref{tab:hyperParemeters}. 

\begin{figure}[t]
    \centering
    \includegraphics[width=.49\textwidth]{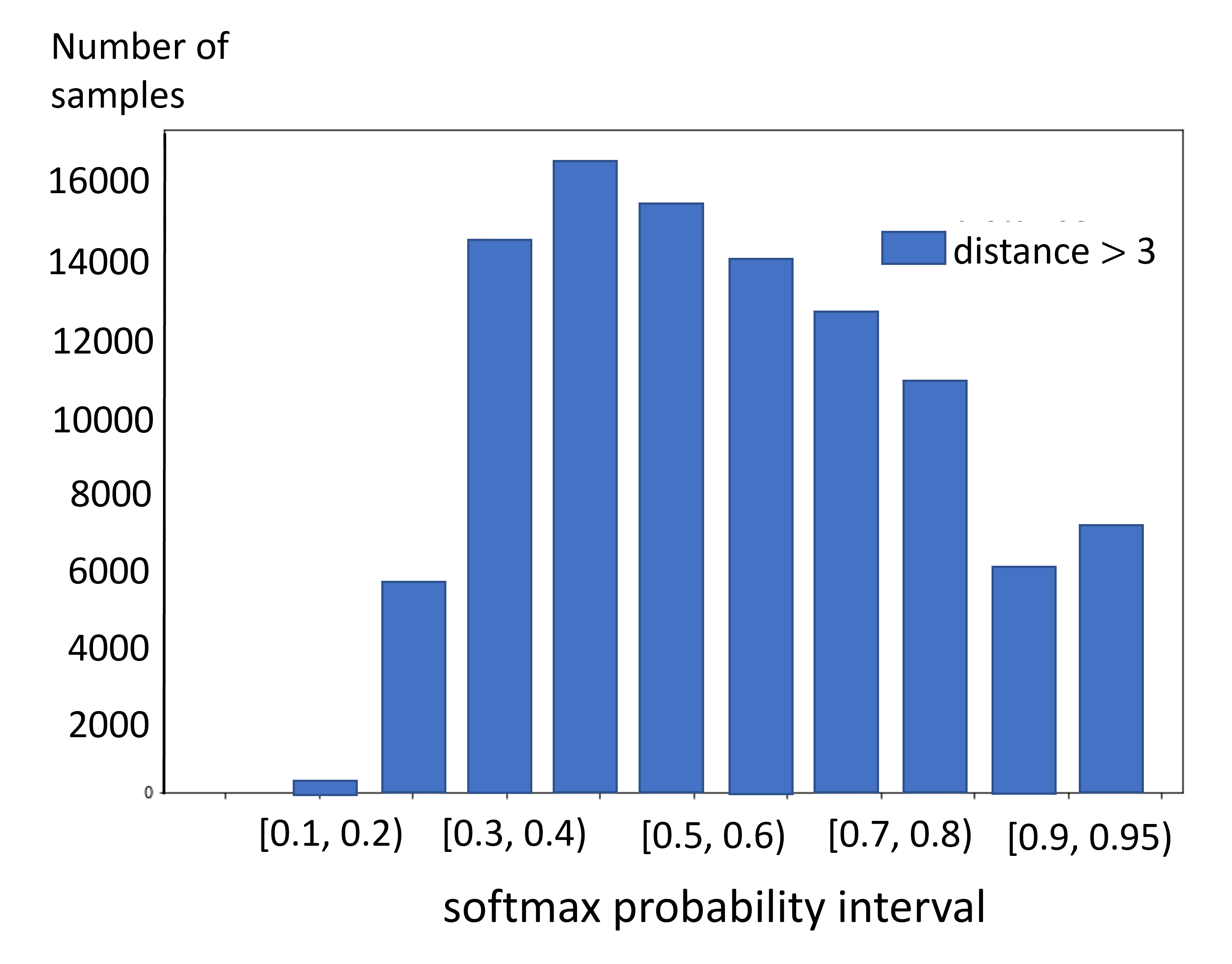}
    %\hfill
    \includegraphics[width=.49\textwidth]{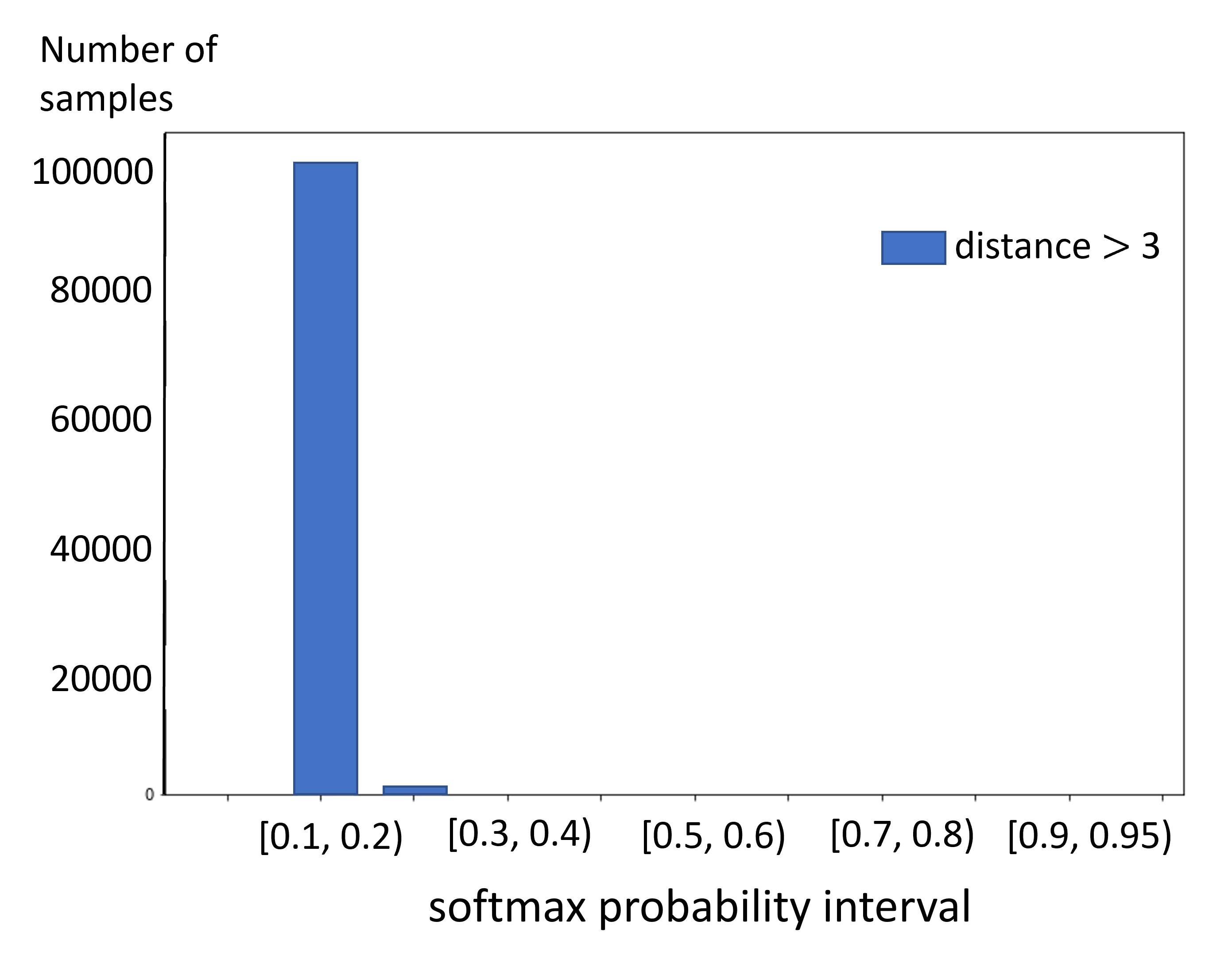}
    \caption{ Statistics of numbers of samples (picked from unsupported corners) per softmax value interval on MNIST (left: before re-training; right: after re-training).}
    \label{fig:postEvalMNIST}
\end{figure}

We first examine if the DNN can output overconfident softmax values for these samples.
From the statistical results, as shown in the left part of Fig.~\ref{fig:postEvalMNIST}, one can find that samples from many unsupported corners (with Hamming distance larger than~$3$ from the training dataset) are assigned a high softmax value. This confirms our conjecture that additional local training is needed to suppress high-confident output against unsupported corners. 
After applying Algorithm~\ref{alg:modification} for fine-tuning the after-monitored-layer sub-network, these unsupported cases are all assigned an averaged softmax value of $\frac{1}{10}$, as shown in the right part of Fig.~\ref{fig:postEvalMNIST}.
Interestingly, the fine-tuning does not deteriorate the accuracy of the neural network on the original training and test sets: We observe a shift from the original accuracy of $99.34\%$ ($98.8\%$) on the training (test) dataset to a new one of $99.24\%$ ($98.84\%$).

\begin{remark}
The repair of the sub-network essentially equips the original network an additional ability of identifying out-of-distribution samples (around the area of unsupported corners) by observing whether the softmax value of prediction is close to $\frac{1}{d_{L}}$ or not.
\end{remark}

\vspace{-2mm}
\subsection{From test case proposal to test case generation}\label{subsec.test.case.gen}

This subsection explores two possibilities for generating inputs that yield features in specific unsupported corners of the monitored layer.

\begin{itemize}
    \item The first method is to verify whether the maximum or minimum activation value of each monitored neuron is responsible for a particular segment or local area of the input, hereafter referred to as Neuron-Wise-Excited-Input-Feature (NWEIF). If such a connection exists, since a corner is a combination of the maximum/minimum activation values of each neuron, then a new input can be formed by combining the NWEIFs of each neuron.

\item The second is to apply optimization techniques. Given an image in the training dataset, perform gradient descent  to find a modification over the image such that the modified image generates a feature within a given unsupported corner.

\end{itemize}

\vspace{-5mm}

\subsubsection{Neuron-wise excited input-feature combination}
We applied the layer-wise relevance propagation (LRP) \cite{montavon2019layer} technique to interpret the images that reach the maximum and minimum five values of a neuron.
LRP is one of the back-propagation techniques used for redistributing neuron activation values at one layer to its precedent layers (possible up to the input layer).
In a nutshell, it explains which parts of the input contribute to the neuron's activation and to what extent.

\begin{figure}[t]
    \centering
    \includegraphics[width=1.05\textwidth]{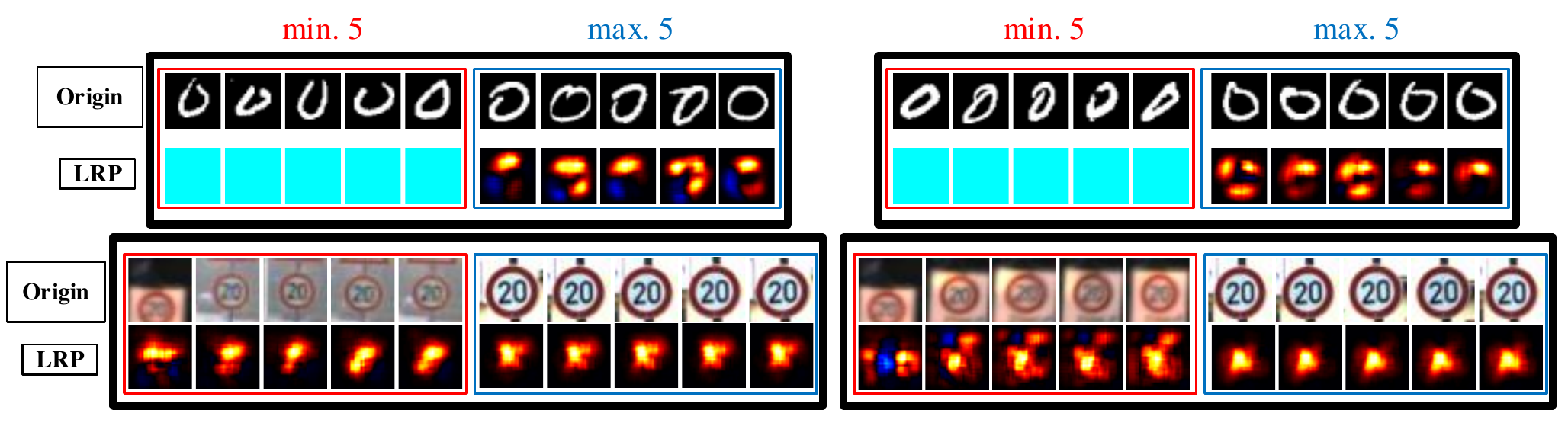}
    \caption{ LRP interpretation example: each bold-black block contains the inputs reaching the minimum 5 (red sub-block) and the maximum 5 (blue sub-block) activation values of a neuron; the top two and the bottom two are from MNIST and GTSRB, respectively.}
    \label{fig:interMNIST}
\end{figure}

%\vspace{-5mm}
\vspace{-2mm}
\paragraph{Discussion} The results in Fig.~\ref{fig:interMNIST} show that it is difficult for humans to compose new inputs based on NWEIF. 
The first and second rows in each bold-black block are the original images and corresponding heat maps interpreted by LRP. 
Although LRP can help us identify regions or features, it is very difficult to precisely associate one neuron with one specific input-feature. We can observe in Fig.~\ref{fig:interMNIST} that for the 20km/h speed sign, the area that leads to maximum activation has considerable overlap with the area that leads to minimum activation. This makes a precise association between neurons and features difficult, justifying the need of using other methods such as optimization-based image generation for testing and Algorithm~\ref{alg:modification} for local training over unsupported corners. 

\subsubsection{Optimization-based test case generation}

\begin{figure}[t]
    \centering
        \includegraphics[width=0.49\textwidth]{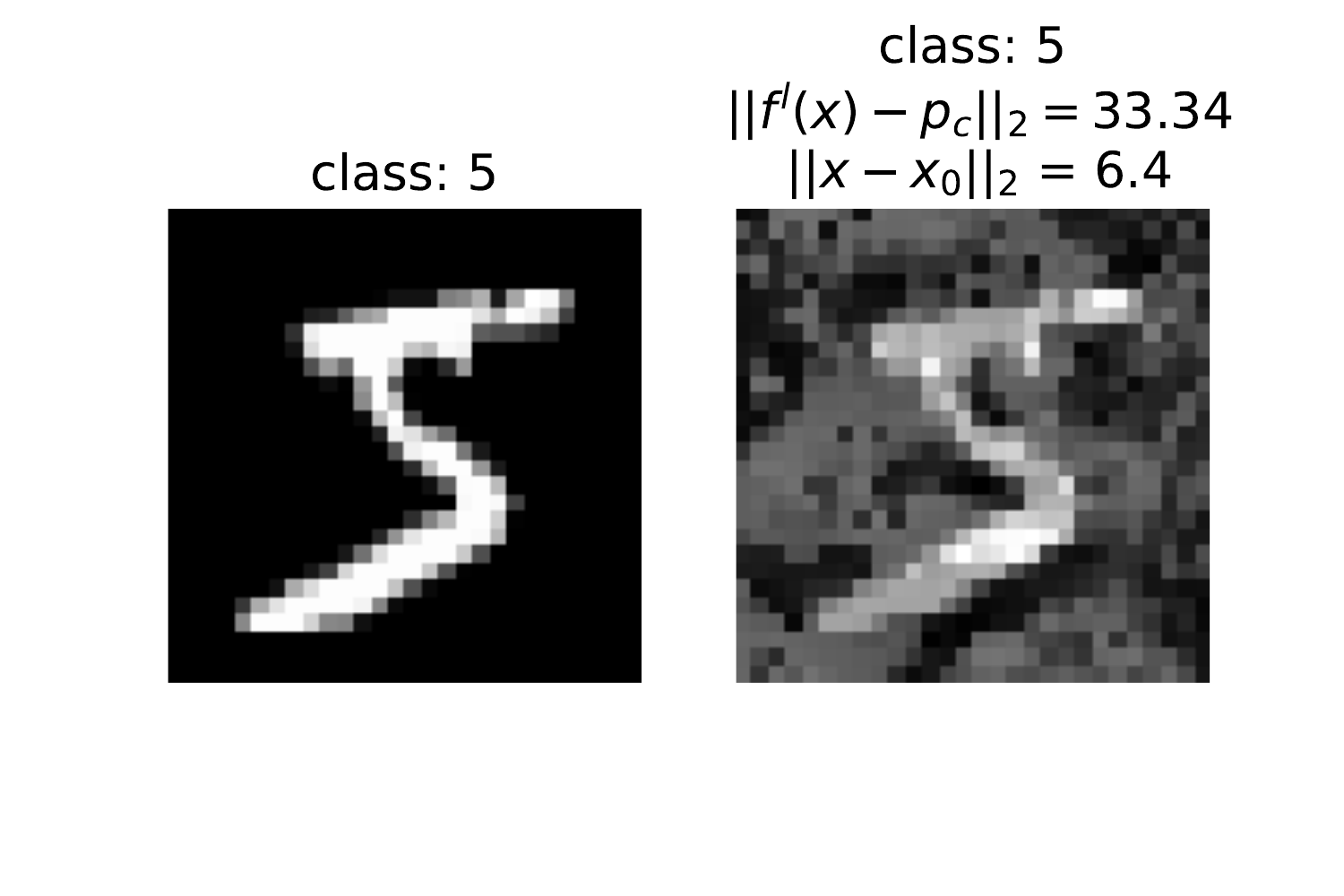}
    \includegraphics[width=0.49\textwidth]{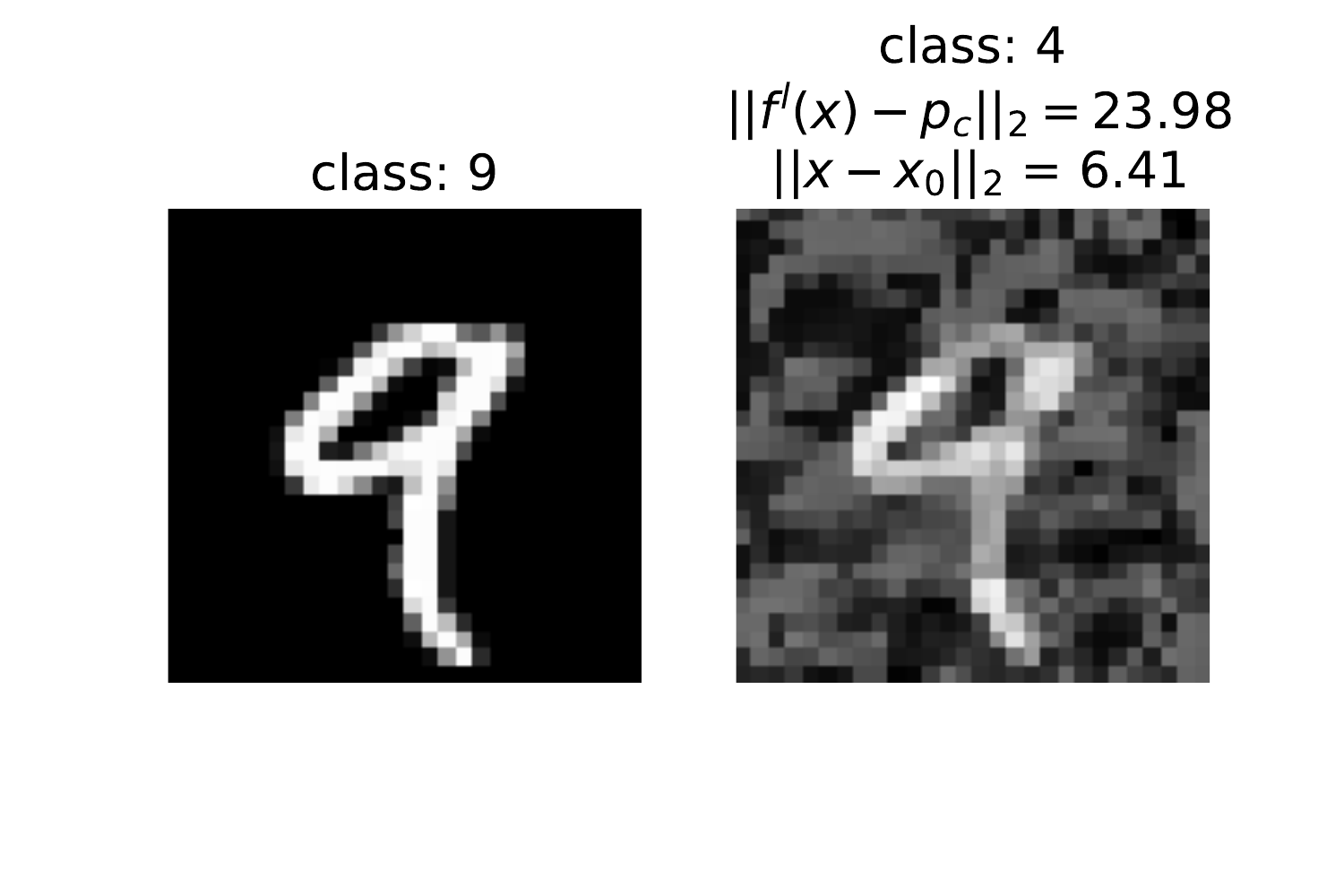}
    
    \vspace{-9mm}
        \includegraphics[width=0.49\textwidth]{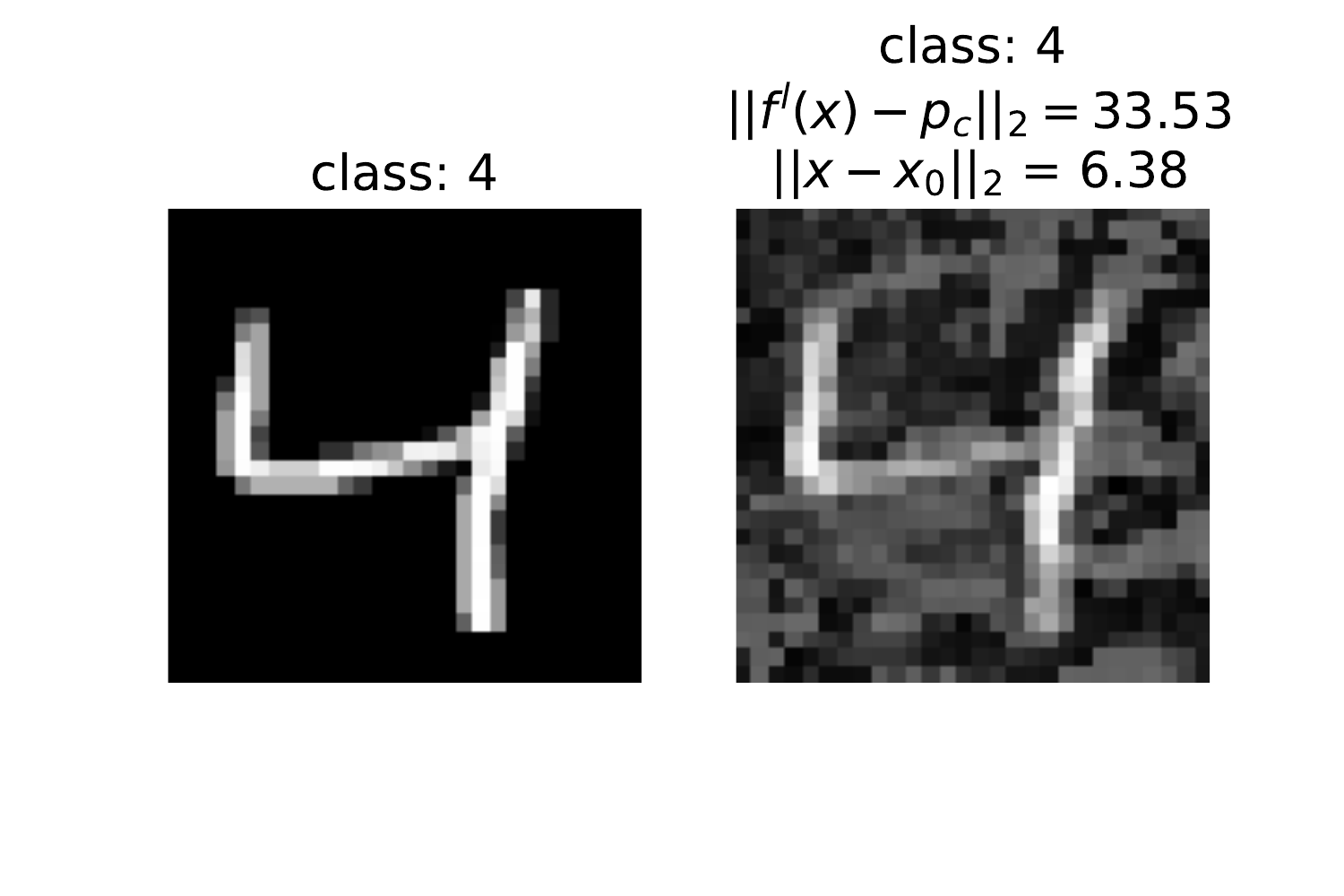}
            \includegraphics[width=0.49\textwidth]{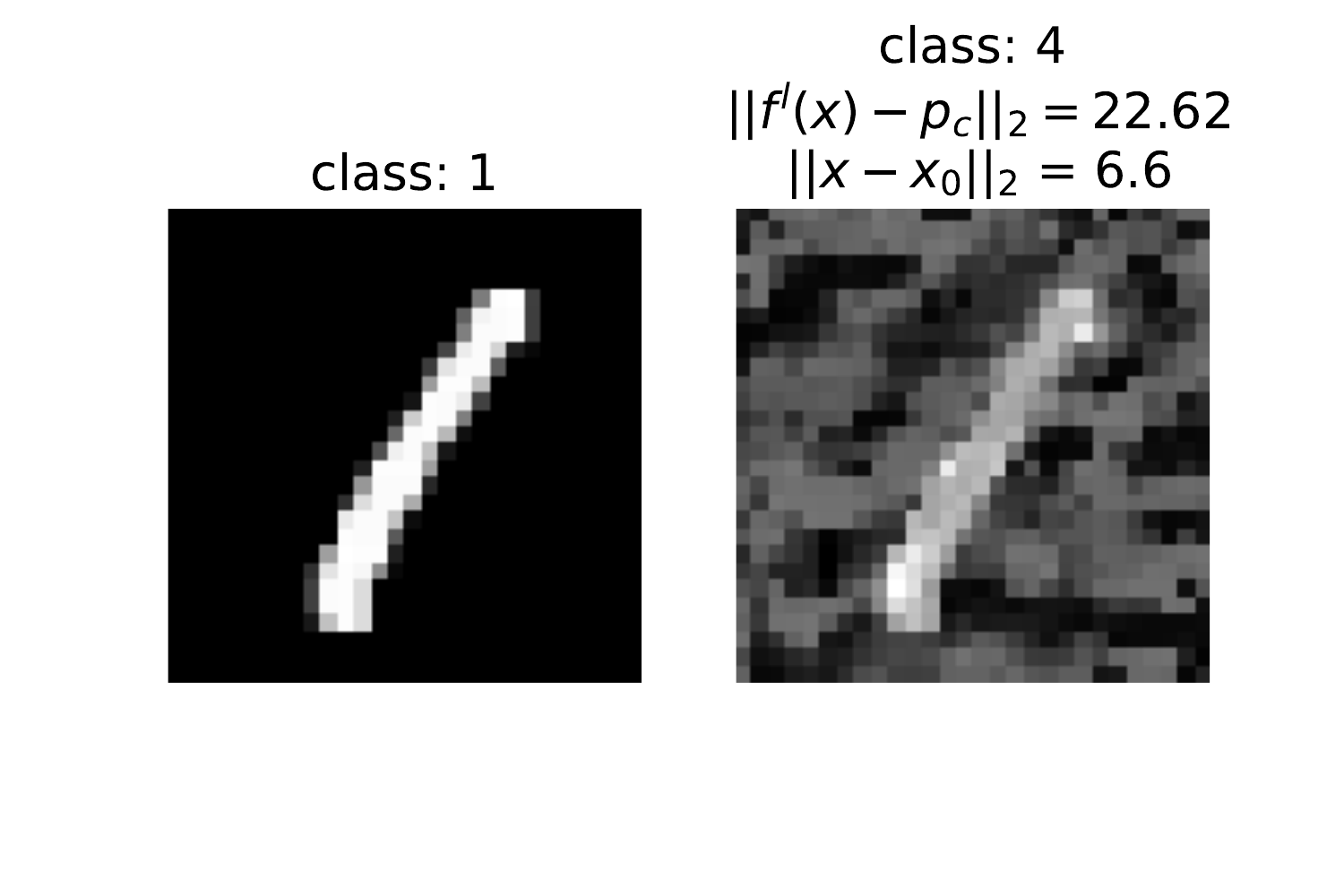}
    \vspace{-10mm}            
            
        \caption{Using adversarial testing to generate images whose feature vectors fall into a particular corner. The original images are shown on the left, and the perturbed ones on the right. We also show the predicted classes, and for the perturbed images additionally the distance of their features from the corner point, as well as the distance from the unperturbed image.}
    \label{fig:advMNIST}
    
    \vspace{-5mm}
\end{figure}

Finally, we create images corresponding to corners by using an optimization method, similar to the ones used for adversarial examples generation \cite{szegedy2013intriguing}. Overall, the generated test case should allow the DNN to (1) fall inside the box of the unsupported corner and to (2) be confident in predicting a wrong class.  In our implementation, the previously mentioned two objectives are integrated as a loss function, which is optimized (by minimizing the loss) with respect to the input image. We refer readers to the appendix for details regarding how such a method is implemented. Figure~\ref{fig:advMNIST} illustrates  examples of original and perturbed images, where for the bottom-right example, the perturbed images not only falls into a particular corner, but the resulting prediction also changes from the initially correct~``$1$'' to the incorrect~``$4$''. 
We observe that when the buffer~$\delta$ around the box is small, it can be difficult for the adversarial testing method to generate images that fall into a specific corner. However, we are unable to state that it is impossible to generate such an input; the problem can only be answered using formal verification. This further justifies the need for local DNN training.

\section{Concluding Remarks}\label{sec:concluding.remarks}
In this paper, we address the issue of testing OoD monitors built from boxed abstractions of feature vectors from the training data, and we show how this testing problem could be reduced to finding corners in hyperrectangle distant from the available training data in the feature space.
The key novelty lies in a rigorous method for analyzing the corners of the monitors and detecting whether a corner is supported or not according to the input in the training data set, generating feature vectors located in the corner.
To the best of our knowledge, it is the first approach for testing the decision boundary of a DNN monitor, where the test prioritization scheme is based on corners (of the monitor) that have no input data being close-by.
The other important result is the DNN repair scheme (via local training) to incorporate the distant-yet-uncovered corners.
To this end, we have developed a tool that provides technical solutions for our OoD detectors based on boxed abstractions. Our experiments show the effectiveness of our method in different applications.

This work raises a new research direction on rigorous engineering of DNN monitors to be used in safety-critical applications. 
An important future direction is the refinement of boxed abstractions: By considering the unrealistic corners, we can refine the abstraction by adding more boxes to remove them.
Another direction is to use some probability estimation method to prioritize corners rather than using Hamming distance.

\vspace{3mm}
\noindent{\textbf{(Acknowledgement)}} This work is funded by the Bavarian Ministry for Economic Affairs, Regional Development and Energy as part of a project to support the thematic development of the Fraunhofer Institute for Cognitive Systems. This work is also supported by the European project Horizon 2020 research and innovation programme under grant agreement No. 956123.

\bibliographystyle{abbrv}
%\bibliography{ref}

\newpage
\section*{Appendix}\label{Appendix}

\section*{A. Optimization-based test case generation}\label{sub.sec.appendix.A}

In this section, we describe our optimization-based method for creating test cases. This method is similar to techniques used for generating adversarial examples in DNNs. 

An adversarial attack tries to perturb an input $\vvector{x}$ so that its classification changes, while simultaneously staying close to the original input. This is achieved by an optimization method with minimizing a loss function that tries to enforce a miss-classification. Apart from that, adversarial attacks also try to ensure that the perturbed input will remain close enough to the original input (since this is the definition of adversarial examples). 

For our case, consider an input $\vvector{x}$ with class $y$, DNN $f$, and let $f^{(l)} (\vvector{x})$ be the output of the $l$-th layer we want to monitor and take features from. Let  $\vvector{p}_c$ be a point in the specified corner $\vvector{c}$ that we want to approximate. We consider the following loss function: $loss(\vvector{x}, y, \vvector{p}_c) = -\lambda \cdot crossentropy( f^{L}\vvector{(x), y)} + ||f^{(l)}(\vvector{x}) - \vvector{p}_c||_2 $. In this loss, the first term is the standard cross-entropy loss for classification, while the second term is the distance of the $l$-th layers features from the corner point $\vvector{p}_c$ with respect $L_2$ norm. The number $\lambda \geq 0$ is a parameter balancing the two objectives. Hence, we can minimize this loss to generate test cases using the Adam optimizer~\cite{kingma2014adam}.

To understand this loss function, note that in a standard adversarial attack only the cross-entropy term is present. The attack tries to minimize the term $-\lambda \cdot crossentropy(\vvector{x}, y)$ (equivalently maximizing the cross-entropy), thus resulting in a miss-classification, e.g. a successful adversarial example.
With our loss now, when it is minimized, we achieve the miss-classification objective, and at the other hand, minimization is achieved by keeping the distance term $ ||f^{(l)}(\vvector{x}) - \vvector{p}_c||_2 $ small. Thus, we achieve a miss-classification, while $f^{(l)}(\vvector{x})$ is close to the corner point~$\vvector{p}_c$, which is what we aim to achieve.

\end{document}